\documentstyle[amssymb,12pt,epsf,epsfig]{article}
\def\beq{\begin{equation}}
\def\eeq{\end{equation}}
\def\bea{\begin{eqnarray}}
\def\eea{\end{eqnarray}}
\def\bq{\begin{quote}}
\def\eq{\end{quote}}

\def\gappeq{\mathrel{\rlap {\raise.5ex\hbox{$>$}}
{\lower.5ex\hbox{$\sim$}}}}
\def\lappeq{\mathrel{\rlap{\raise.5ex\hbox{$<$}}
{\lower.5ex\hbox{$\sim$}}}}

\begin{document}

\pagestyle{empty} 
\begin{flushright}
{CERN-TH/2002-241}
\end{flushright}
\vspace*{5mm}

\begin{center}
{\bf QUARK MASSES FROM QUARK--GLUON CONDENSATES IN A MODIFIED PERTURBATIVE QCD
} \\[0pt]
\vspace*{1cm} {\bf A. Cabo}\\[0pt]
\vspace{0.3cm} Theoretical Physics Division, CERN \\[0pt]
CH - 1211 Geneva 23, Switzerland \\[0pt]
\vspace*{2cm} {\bf ABSTRACT} \\[0pt]
\end{center}

\vspace*{5mm} \noindent\ \ \ \ \ \ \ \ \ \ \ \ \ \ \ \ \ \ \ \ \ \ \ \ \ \ \
\ \ \ \ \ \ \ \ \ \ \ \ \ \ \ \ \ \ \ \ \ \ \ \ \ \ \ \ \ \ \ \ \ \ \ \ \ \
\ \ \ \ \ \ \ \ \ \ \ \ \ \ \ \ \ \ \ \ \ \ \ \ \ \ \ \ \ \ \ \ \ \ \ \ \ \
\ \ \ \ \ \ \ \ \ \ \ \ \ \ \ \ \ \ \ \ \ \ \ \ \ \ \ \ \ \ \ \ \ \ \ \ \ \
\ \ \ \ \ \ \ \ \ \ \ \ \ \ \ \ \ \ \ \ \ \ \ \ \ \ \ \ \ \ \ \ \ \ \ \ \ \
\\
In this note, it is argued that the mass matrix for the six quarks can be
generated in first approximation by introducing fermion
condensates on the same lines as was done before for gluons, within the
modified perturbative expansion for QCD proposed in former works. Thus,
the results point in the direction of the conjectured link of the
approximate `Democratic' symmetry of the quark mass matrix and
`gap' effects similar to the ones occuring in superconductivity. The
condensates are introduced here non-dynamically and therefore the
question of the possibility for their spontaneous generation remains open.
However, possible ways out of the predicted lack of the `Democratic'
symmetry of the condensates resulting from the
spontaneous breaking of the flavour symmetry are suggested. They come from an
analysis based on the Cornwall--Jackiw--Tomboulis (CJT)
effective potential for composite operators.

\vspace*{2cm}

\noindent

\begin{flushleft} 

\vspace*{1cm}
CERN-TH/2002-241 \\
September 2002\\
\vspace*{1cm}
$*$ Permanent address: Group of Theoretical Physics, Instituto de Cibern\'etica, 
Matem\'atica y F\'{\i}sica, Calle E, No. 309, Vedado, La Habana, Cuba. 
Email: cabo@cidet.icmf.inf.cu. 
\end{flushleft}
\vfill\eject

\setcounter{page}{1} \pagestyle{plain}


\section{Introduction}

\ The nature of the chiral and flavour symmetries in QCD and the
Standard Model is one of the open fundamental questions in high 
energy physics \cite{leut,fritzsch,fritzsch1,lane,colwitt,vafwitt}.
The amount of experimental
information in direct relation with chiral and flavour physics that remains
to be properly understood is enormous. This situation, implies
that these problems deserve to be considered along many directions in 
search of a better understanding.

\ In previous works we studied the consequences of modifying the Feynman
rules of the perturbative QCD in a way that incorporates the presence of a
condensate of zero-momenta gluons in the initial state for constructing
the Wick expansion \cite{mpla, prd,epjc,hoyer}. In \cite{prd} the modified
theory included only a new term in the gluon propagator. This form of
the change resulted from the adiabatic connection of the interaction on a modified
vacuum of the non-interacting theory. The new state was chosen in a form similar 
to the way the BCS wavefunction of superconductivity is sometimes represented, i.e. 
as an exponential of a quadratic polynomial in the creation operators. For the BCS
wavefunction two electron creation operators are employed. In our discussion
a quadratic polynomial in the gluon and ghosts creation operators were
considered. The colourless character and the additional requirement that the
new vacuum be a physical state of the free theory in the
non-interacting limit, strictly defined the form of the polynomial. Also, the
Lorentz invariance requirement led us to consider only creation
operators near the zero momentum limit.

\ The above propagator, which was employed in a simple tree calculation
directly led to a value of the gluon condensate
parameter (defined as the mean value of the gauge Lagrangian) 
different from zero. Moreover,
when fixing the only parameter $C$ 
to reproduce a current estimate for $\langle g^{2}G^{2}\rangle$ the one-loop 
corrections for the quark masses surprisingly
gave as the result one third of the nucleon mass value \cite{epjc}. \ 

\ The mentioned elements naturally suggested the following idea: if the 
BCS-like modification of the gluon state could lead to such interesting conclusions,
as the prediction of the constituent masses for quarks, it could be
reasonable to expected that a similar state for the quarks, introduced in
the massless QCD, could describe the Lagrangian quark masses. This
is the basic question addressed in this note. 

\ The answer, at least in this preliminary stage, is positive. After modifying
the free-quark propagator by introducing the zero momentum terms
representing the analogue to Cooper pair condensate in this problem (that
is, all the colourless and uncharged quark--antiquark condensates) the
simplest approximation for the Dyson equation for quarks produces a diagonal
Lagrangian mass matrix by the simple choice of a structure  also diagonal, 
for the quark condensates. A second possibility is also analysed it
corresponds to the case in which the `Democratic' symmetry proposed by H.
Fritzsch \cite{fritzsch,fritzsch1}, is assumed for the condensate matrix.
Then, it turns out that this symmetry is translated to the quark mass matrix
and the spectrum produces two massive quarks each having  two
massless counterparts at the level of approximation considered. That is, the
six initially equivalent quarks are decomposed in two triplets resembling
the $u$ and $d$ type sets $(u,c,t)$ \ and $\ (d,s,b).$

\ Further, the inclusion of the zero momentum piece of the gluon propagator
reflecting the gluon condensate is also examined in order to check for
consistency with the results obtained in \cite{epjc} for the constituent
masses when the Lagrangian quark masses were taken as parameters. 
The results for
the six quarks were qualitatively in correspondence with the former
evaluation. Again, the light quark masses turned out to have near by 1/3 of the
proton mass and the massive ones practically coincided with their
Lagrangian values. Thus, the picture seems to be consistent, up to these
first approximations. \ 

\ Finally, comments are given about the possibility of justifying the
pattern of quark condensates reproducing the quark masses, as generated by a
dynamical breaking of the chiral symmetry. Possible solutions are suggested
of the problem related with the indications of a necessary
equality of the dynamically generated fermions condensates \cite{colwitt,
lane}. The basic argument given here consists in first underlining the relevance, for
the determination of the form of the chiral symmetry breaking, that have the gap
equations considered in the CJT effective action approach for composite
operators; then, and more importantly, it is  to indicate the possible need of the
inclusion of  more than one loop fermion corrections for the CJT effective
action for the quark and gluon propagators. These terms are directly
connected with the existence of attractive channels between different
fermions and should be of relevance in determining the strength of the
condensate. They are linked with the two-particle Bethe--Salpeter propagator
whose associated bound state equation is in turn in close correspondence with the
existence and strength of the condensate of Cooper pairs in the BCS
description of superconductivity. On the other hand, the argument implying
the equality of all the condensates rests on what could be at first sight
the reasonable assumption, that for a large number of colours, the one 
loop correction to the effective action should be the leading 
contribution \cite{colwitt}.

\ The paper is organized as follows: in Section 2, the corrections
to the one loop quark self-energy  and the related dispersion equations 
are discussed.  Section 3 is devoted to checking the
effect of the additional consideration of the gluon condensate term in the
propagator. Finally, in this same final section, possible ways are suggested 
through which non-equal condensates for all the fermions could emerge dynamically.

\section{Lagrangian quark masses in terms of condensates}

\ As mentioned in the introduction, we will consider within the first
approximation, the effects of employing a modified quark propagator
reflecting the presence of condensates of quarks in the vacuum of
the free massless QCD. The specific form of the propagator will be 
\[
G_{q}^{i_{1}i_{2};f_{1}f_{2}}(p)=\left( -\frac{\gamma ^{\mu }p_{\mu }\ \delta
^{^{f_{1}f_{2}}}}{p^{2}+i\epsilon }+i\ \delta (p)\ C^{f_{1}f_{2}} \right)\ \delta
^{i_{1}i_{2}},\ \,\,i,j=1,2,3,
\]
where the structure has a diagonal form in the colour indices, and the
real and symmetric matrix in the flavour indices $C^{f_{1}f_{2}}$ reflects
the consideration of all the possible sorts of \ colourless quark--antiquark
condensates. As shown in Refs. \cite{prd,hoyer} \ this zero momentum
contribution for fermions can be obtained after considering as the initial state,
before the connection of the interaction, a modified vacuum in the form of
an exponential of quadratic products of quark and antiquark creation
operators as 
\begin{equation}
\mid 0 \rangle =\lim_{p^{\ast }\rightarrow 0}\exp \left( \sum_{f_{1}f_{2}}%
\tilde{C}_{q}^{f_{1}f_{2}}(p^{\ast })\ \bar{q}_{f_{1}}^{+}\ (p)\
q_{f_{2}}^{+}(p^{\ast })\right)   \label{condstat}
\end{equation}
in which the limit of a small null momentum $p^{\ast }\rightarrow 0 $ should
be taken in the corresponding infinite volume limit $V\backsim 1/\mid 
\overrightarrow{p^{\ast }}\mid ^{3} \rightarrow \infty $ , where $%
\overrightarrow{p^{\ast }}$ is the spatial part of $p^{\ast }.$ The 
structure of this state is similar to the one employed for the representation of the
BCS wavefunction, and the attractive and very strong character of the colour
forces, at least naively suggest the relevance of this approach to the
description of QCD. \ Even in the presence of the weak phonon attraction with respect
with the strong Coulomb repulsion between electrons, \ the Cooper
pairs condensate becomes essential for the description of the
superconducting state. Then, it is hard to think that in the case of the
attractive and even confining attraction due to colour interaction the vacuum
without condensates could be stable.

\ In order to consider the first implications of the introduced modification,
lets us disregard for the moment the gluon condensate and evaluate the
contribution of the change in the quark propagator to the one loop self-energy,
associated to the first diagram in Fig. 1. \ The quark and gluon lines with central 
dots represent the
added terms in the quark and gluon  propagators. \ As the loop integration is
annihilated by the delta function at zero momentum, the result is simply
given by
\begin{eqnarray}
\Sigma _{q}^{i_{1}i_{2};f_{1}f_{2}}(p) &=&\frac{\gamma _{\mu }\gamma ^{\mu
}g^{2}}{(2\pi )^{4}}\ T_{i_{1}l}^{a}\frac{C^{f_{1}f_{2}}}{p^{2}}\ T_{l\
i_{2}}^{a}  \nonumber \\
&=&\frac{4g^{2}}{(2\pi )^{4}}\ T_{i_{1}l}^{a}\frac{C^{f_{1}f_{2}}}{p^{2}}\
T_{l\ i_{2}}^{a}  \nonumber \\
&=&\frac{4g^{2}C_{F}}{(2\pi )^{4}}\ \frac{C^{f_{1}f_{2}}}{p^{2}}\ .
\label{self1}
\end{eqnarray}
where use has been done of the relations
\begin{eqnarray*}
(\gamma _{\mu }p^{\mu })^{2} &=&p^{2}, \\
T_{i_{1}l}^{a}\ T_{l\ i_{2}}^{a} &=&C_{F}\ \delta ^{i_{1}i_{2}}\  \\
C_{F} &=&\frac{N^{2}-1}{2N}.
\end{eqnarray*}
\ Note that a better procedure could be to sum up all the mass insertions
in the internal quark propagator joining the two vertices. However, here we
are mainly interested in considering the simplest approximation. It should be
noticed that this term can be viewed, in conjunction with its gluonic 
counterpart in Fig. 1 (to be
introduced in the next section for treating gluon condensation), as the
only two terms in a sort of `modified tree' approximation for the self-energy in
which no `modified loop' radiative corrections are considered. The search
for a systematic formulation for an expansion of this kind will
be considered elsewhere. The relevance of such loop expansion in
determining an alternative and more rapidly convergent perturbative scheme
is also suggested, for example, by the similarly existing procedures for
Bose condensation phenomena. There, the Bogoliubov shift in the amplitude
of the scalar condensate field to the minimum, leads to a satisfactory
perturbation expansion in terms of the new modified radiative field.


\begin{figure}
\begin{center}
\epsfig{figure=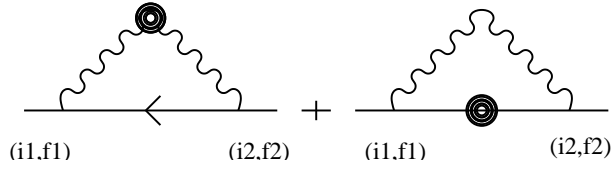,width=8cm}
\end{center}
\caption{The condensate contributions to the one loop selfenergies. Note
that the zero momentum Dirac delta functions anihilate the only existing 
integration. This circumstance suggests the existence of a systematic
`modified loop' approximation in the problem.}
\label{Fig.2}
\end{figure}

 \ After substituting the self-energy approximate expression (\ref{self1}) in
the Dyson equation for the propagator, and considering the quark wave modes
solving the homogeneous version of the equation, it follows
\begin{eqnarray}
0 &=&\left( -p_{\mu }\gamma ^{\mu }\ \delta ^{i_{1}i_{2}}\ \delta
^{f_{1}f_{2}}-\Sigma _{q}^{i_{1}i_{2};f_{1}f_{2}}(p)\right) \Psi
_{f_{2}}^{i_{2}}(p)  \label{dyson} \\
&=&\delta ^{i_{1}i_{2}}\left( -p_{\mu }\gamma ^{\mu }\ \ \delta
^{f_{1}f_{2}}-\frac{4g^{2}C_{F}}{(2\pi )^{4}}\ \frac{C^{f_{1}f_{2}}}{p^{2}}%
\right) \Psi _{f_{2}}^{i_{2}}(p)  \nonumber \\
&=&\frac{1}{p^{2}}\left( -p^{2}p_{\mu }\gamma ^{\mu }\ \ 
\delta^{f_{1}f_{2}}-\frac{4g^{2}C_{F}}{(2\pi )^{4}}\ C^{f_{1}f_{2}}
\right) \Psi _{f_{2}}^{i_{1}}(p),  \nonumber
\end{eqnarray}
which after being multiplied by the matrix between the parentheses with the
second term having opposite sign, results in
\begin{equation}
\left( (p^{2})^{3}\ \ \delta ^{f_{1}f_{2}}-\left( \frac{g^{2}C_{F}}{4\pi ^{4}}
\right)^{2}\ (C^{2})^{f_{1}f_{2}}\right) \Psi _{f_{2}}(p)=0.  \label{dispersion}
\end{equation}

Therefore, it can be noticed that, within the
considered approximation, the quark masses are defined by the eigenvalues of
the square of the matrix of the condensates. Below,
two phenomenological possibilities for fixing the condensate matrix
will be treated.

\subsubsection{ Diagonal condensate matrix}

This case corresponds to directly selecting $C_{^{f_{1}f_{2}}}$ as a diagonal matrix in the
flavour indices in order that its square multiplied by $(\frac{g^{2}C_{F}}{
4\pi ^{4}})^{2}\ $, substituted in (\ref{dispersion}) will produce 
the mass matrix for the six quarks. \ Thus $C_{^{f_{1}f_{2}}}$ takes the form
\begin{equation}
g^{2}C_{^{f_{1}f_{2}}}\equiv \frac{4\pi ^{4}}{C_{F}}\left[ 
\begin{array}{cccccc}
m_{u}^{3} & 0 & 0 & 0 & 0 & 0 \\ 
0 & m_{d}^{3} & 0 & 0 & 0 & 0 \\ 
0 & 0 & m_{s}^{3} & 0 & 0 & 0 \\ 
0 & 0 & 0 & m_{c}^{3} & 0 & 0 \\ 
0 & 0 & 0 & 0 & m_{b}^{3} & 0 \\ 
0 & 0 & 0 & 0 & 0 & m_{t}^{3}
\end{array}
\right] ,  \label{matrizC}
\end{equation}
which, after using the link between this matrix and the condensate values
\begin{equation}
\langle 0 \mid \Psi _{f_{1}}^{i}(0)\overline{\Psi _{f_{2}}^{i}}(0)^{i}\
\mid 0 \rangle =\frac{4N}{(2\pi )^{4}}C^{f_{1}f_{2}},  \label{masscond}
\end{equation}
leads to the following connection between the chiral condensates and the
corresponding mass of the quark of flavour $f:$
\begin{equation}
g^{2}\langle 0\mid \Psi _{f}^{i}(0)\overline{\Psi _{f}^{i}}(0)^{i}\ \mid 0\
\ \rangle =\frac{2(N^{2}-1)}{N}\ m_{f}^{3},\ \,\,\ f=1,...,6.
\end{equation}

 The compatibility of these relations between quark condensates 
and masses with the existing 
extensive experimental and numerical information will be 
considered elsewhere.

\subsubsection{ `Democratic' condensate matrix}

\ The other selection to be analysed is the one which will produce as a 
result the `Democratic' symmetry (which relevance was argued in\ \cite
{fritzsch}) for each one of two separate quark triplets that we will call
$u$ and $d$ type quarks. As has been argued\ \cite{fritzsch},
assuming that this first stage of the symmetry breaking can be justified
dynamically, then the masses for the other massless modes can be produced
at next steps of approximation. Thus, it has some interest
to verify that such a structure of the mass matrix can also be
generated by the presence of quark condensates. The matrix $C_{f_{1}^{\ast
}f_{2}^{\ast }}$, showing the symmetry in the above mentioned
sectors, takes the form 
\[
g^{2}C_{f_{1}^{\ast }f_{2}^{\ast }}\equiv \frac{4\pi ^{4}}{3C_{F}}\left[ 
\begin{array}{cccccc}
m_{b}^{3} & m_{b}^{3} & m_{b}^{3} & 0 & 0 & 0 \\ 
m_{b}^{3} & m_{b}^{3} & m_{b}^{3} & 0 & 0 & 0 \\ 
m_{b}^{3} & m_{b}^{3} & m_{b}^{3} & 0 & 0 & 0 \\ 
0 & 0 & 0 & m_{t}^{3} & m_{t}^{3} & m_{t}^{3} \\ 
0 & 0 & 0 & m_{t}^{3} & m_{t}^{3} & m_{t}^{3} \\ 
0 & 0 & 0 & m_{t}^{3} & m_{t}^{3} & m_{t}^{3}
\end{array}
\right], 
\]
where the components of the flavour index vector used up to now, 
$f=u,d,s,c,b,t$, \ have been permuted to define the new flavour index
$f^{\ast }=d,s,b,u,c,t.$  This change  was made in order render the
notation closer to the one employed in the literature for discussing these matrices.
It can be noticed that the square, and even any kind of powers, of this
matrix also retains the same structure modulo  constant factors. Thus, the
mass matrix of the quarks, which can be obtained as being proportional to
the power $1/3$ of the matrix $C_{f_{1}^{\ast }f_{2}^{\ast }}$, also
have the same form. Henceforth, the quark mass matrices proposed \ in \cite
{fritzsch} can also be generated by the presence of appropriate condensates
in the vacuum having the same symmetry.

The matrix $C_{f_{1}^{\ast }f_{2}^{\ast }}\ $above can be diagonalized by
a unitary transformation $U$\ to produce, for the quark mass matrix :
\[
M_{f_{1}^{\ast }f_{2}^{\ast }}=(\frac{g^{2}C_{F}}{4\pi ^{4}})^{\frac{1}{3}%
}(U^{-1}C^{\frac{1}{3}}U)_{f_{1}^{\ast }f_{2}^{\ast }}\equiv \left[ 
\begin{array}{cccccc}
0 & 0 & 0 & 0 & 0 & 0 \\ 
0 & 0 & 0 & 0 & 0 & 0 \\ 
0 & 0 & m_{b} & 0 & 0 & 0 \\ 
0 & 0 & 0 & 0 & 0 & 0 \\ 
0 & 0 & 0 & 0 & 0 & 0 \\ 
0 & 0 & 0 & 0 & 0 & m_{t}
\end{array}
\right] . 
\]

\ Thus, the first steps in the approach related with the approximate
`Democratic' symmetry of the experimental quark mass spectrum is also implied by the
presence of the chiral condensates with the same symmetry as treated
in the modified perturbative expansion. 

\section{Constituent masses from the gluon condensation}

\ Up to now, for the sake of clarity in this first
exploration, we have considered only the new element with respect to the
gluon condensate effects studied before. \ Let us now examine how the
Lagrangian quark masses, previously fixed in terms of the quark condensates,
are modified by the introduction of the gluon condensate. The ideal situation
would be that the results in this first step could reproduce 
the spectrum of masses obtained in \cite{epjc} qualitatively. We will see 
that in effect this is what happens

\ The \ modified propagator for the gluon has the form \ \ 
\begin{equation}
G_{\mu \nu }^{ab}(p)=\left( \frac{1}{p^{2}+i\epsilon }-i\ \delta (p)\ C \right)\ \delta
^{ab}g^{\mu \nu },  \label{propglue}
\end{equation}
which, after being inserted in the expression associated to the second diagram
in Fig. 1 for the quark self-energy and integrated over the loop momentum,
gives 
\begin{equation}
\Sigma _{g}^{i_{1}i_{2};f_{1}f_{2}}(p)=-\frac{M^{2}\ \delta
^{i_{1}i_{2}}\delta ^{f_{1}f_{2}}p_{\mu }\gamma ^{\mu }}{p^{2}},
\label{self2}
\end{equation}
where the value of the constant $C$ determined in \cite
{epjc} from reproducing a current estimate for the gluon condensate
parameter $\langle g^{2}G^{2} \rangle$ has been used. The constants 
that appear have the values 
\begin{eqnarray}
M &=&\sqrt[2]{\frac{2C_{F}\ g^{2}C}{(2\pi )^{4}}=}333\ \,\,MeV/c^{2},
\label{ctc} \\
g^{2}C &=&65\ \,\, (GeV/c^{2})^{2}.  \nonumber
\end{eqnarray}

Substituting the two contributions to the self-energy in the Dyson equation,
the quark modes then satisfy
\begin{equation}
\frac{1}{p^{2}}\left( -p^{2}p_{\mu }\gamma ^{\mu }\left( 1-\frac{M^{2}}{p^{2}}
\right)\ \
\delta ^{f_{1}f_{2}}-\frac{4g^{2}C_{F}}{(2\pi )^{4}}\ C^{f_{1}f_{2}}\right)
\Psi _{i}^{f_{2}}(p)=0,  \label{dysonglue}
\end{equation}
which gives, again by taking the matrix obtained by changing the relative sign
of the two terms in between the parentheses, and multiplying the equation by
the result:
\begin{equation}
\frac{1}{p^{2}}\left( (p^{2})^{3}\ \left( 1-\frac{M^{2}}{p^{2}} 
\right)^{2}\ -(\frac{
g^{2}C_{F}}{4\pi ^{4}})^{2}\ (C_{f}^{2})\right) \Psi _{i}^{f}(p)=0,
\label{disperglue}
\end{equation}
where it has been assumed that the quark modes are already diagonalizing
the matrix $C^{f_{1}f_{2}}$ with eigenvalues $C_{f}$ for each flavour $f.$

It can be noticed that the quark mass matrix is again determined by the
square of the matrix of the quark condensates, but the\ new dispersion
relations for the different quark polarizations associated to the
eigenvalues $C_{f}$ \ become
\begin{equation}
p^{2}\ (p^{2}-M^{2})^{2}\ \delta \ ^{f_{1}f_{2}}-\left(\frac{g^{2}C_{F}}{4\pi ^{4}
} \right)^{2}\ C_{f}^{2}=0.  \label{dysonqg}
\end{equation}

Let us consider the resulting mass spectrum under the following
assumptions: a) the quark condensate parameters are fixed by determining the
Lagrangian mass matrix, as was done in the previous section, and \ b) the $%
M$ constant has been fixed as before by reproducing the value of the gluon
condensate.  Then the solution of the Dyson equation (\ref{dysonqg})
determines the following results for the pole quark masses:
\begin{center}
\begin{tabular}{|c|c|c|c|}
\hline
$Quark\ q$ & $m_{qLow}^{Exp}(MeV)$ & $m_{qUp}^{Exp}(MeV)$ & $%
m_{q}^{Theo}(MeV)$ \\ \hline
$u$ & $1.5$ & 5 & 333--0 \\ \hline
$d$ & $3$ & 9 & 333--0 \\ \hline
$s$ & $60$ & 170 & 339--326--13 \\ \hline
$c$ & 1100 & 1400 & 1255 \\ \hline
$b$ & 4100 & 4400 & 4233 \\ \hline
$t$ & 168600 & 17900 & 173500 \\ \hline
\end{tabular}
\end{center}
where the second and third columns give the top and bottom experimental
bounds of the quark Lagrangian masses. As can be observed from the data, 
the light quarks $u$, $d$ and $s$ got
values close to 1/3 of the nucleon mass, as was also the outcome in \cite
{epjc}. Similarly also, the massive quark masses  have remained
almost invariant under the addition of the gluon condensate. The next salient
feature of the results in the table is the fact that the strange quark
(making honour to his name!) had three solutions at this approximation. One
of them has the very low mass of 13 MeV, and the other two are  in the same
region as the value of the common mass (333 MeV) for the $u$ and $d$ quarks,
but one lies over and the other below. \ It could be the case that, after
introducing the above mentioned sort of \ `modified' one loop
approximation, the results can tend to approach even more those \
obtained in \cite{epjc}, where the Lagrangian quark masses were introduced
explicitly.

Finally, it is needed to remark that various almost massless modes have been
obtained. This outcome should be expected in the case when the symmetry has been
spontaneously broken. Therefore, the appearance of massless modes is, at least,
not in contradiction with the spontaneous generation of the quark
condensates from the vacuum in a similar way, as the gluon condensate seems to be
generated from the vacuum in the first approximations \cite{mpla}, closely 
resembling the picture in the early Savvidi's chromomagnetic field approach. 

\subsubsection{Comment on the spontaneous generation of the quark condensates}

Finally, the question of the spontaneous generation of the condensates from
the vacuum will be analysed. Let us consider for concreteness the effective
potential for composites operators introduced in  \cite{cjt} as applied to
the massless QCD. This functional can be written in the following form
\begin{eqnarray}
V[A,\Psi \Psi ,G_{g},G_{q}]_{(A,\Psi \Psi =0)} &=&\frac{i}{2} Tr [\log
[G_{g}^{(0)-1}G_{g}-G_{g}^{(0)-1}G_{g}]+  \label{veff} \\
&&-i Tr[\log [G_{q}^{(0)-1}G_{q}-G_{q}^{(0)-1}G_{q}]+  \nonumber \\
&&V^{(2)}[A,\Psi \Psi ,G_{g},G_{q}]_{(A,\Psi \Psi =0)},\nonumber
\end{eqnarray}
where $V$ depends on arbitrary values of the mean quantum fields
(here the quark and gluon ones) and $G_{g},G_{q}$ are also arbitrarily fixed
expressions for their propagators. \ The $V^{(2)}$ functional represents
contributions higher than one loop  in terms of Feynman diagrams. The mean
field values will be assumed to vanish in the ground state as required by
Lorentz invariance, and they will be omitted below. The free
inverse propagators as considered in the previous section correspond to the
massless QCD and are given by 
\begin{eqnarray*}
G_{q}^{(0)-1} &\equiv &-p_{\mu }\gamma ^{\mu }, q=(u,d,s,c,b,t) \\
G_{g}^{(0)-1} &\equiv &G_{g,\mu \nu }^{(0)-1ab}(p)=p^{2}\ \delta ^{ab}g^{\mu
\nu }.
\end{eqnarray*}

The traces in the fist term in (\ref{veff}) are in the spinor, fundamental 
colour and spatial indices and for the second term they are 
over the Lorentz, adjoint colour and spatial indices of the operators.

In order to analyse the spontaneous
generation of the condensates from the vacuum, the propagator
candidates showing non-vanishing condensates should be substituted in (\ref
{veff}) in order to verify if they have or not lower potential than the ones
showing zero values for the condensates. It could be helpful to recall that
this should be the case, because the effective potential satisfies the
condition of giving the minimum mean values of the energy in the subspace of
the states of the system satisfying the constraints of having fixed the mean
fields and propagators (which also are mean values).

 It is interesting to notice that in the really first approximation for the
propagators, in which only the delta function modifications are included,
the first two terms in the above expression for the effective potential are
completely independent\ of the values of condensate parameters. This is
simply because the modifications introduced are also allowed free propagators
for the theory, only differing from the Feynman ones in the rounding of
the singularity. \ However, even at the simplest one-loop approximation,
the propagator modes get masses and the mentioned two terms of the effective
potential will depend on the selected condensation parameters. \ 

The gluon and quark propagators, in the approximation in which the
self-energies are defined by (\ref{self2}),(\ref{self1}), take the form 
\begin{eqnarray*}
G_{g\mu \nu }^{ab}(p) &=&\frac{1}{(p^{2}+m_{g}^{2})}\ \delta ^{ab}g^{\mu \nu
}, \\
G_{q}(p) &=&\frac{\delta ^{f_{1}f_{2}}}{\left( -p_{\mu }\gamma ^{\mu }(1-%
\frac{M^{2}}{p^{2}})\ \ -\frac{S_{f_{1}}}{p^{2}}\ \right) },
\end{eqnarray*}
where the new parameters, which simplify the notation but also characterize
the gluon and quark condensates, are related with the already
defined constants $C$ and $C_{f},(f=u,d,s,c,b,t)$ through
\begin{eqnarray*}
m_{g}^{2} &=&\frac{6g^{2}C}{(2\pi )^{4}}, \\
S_{f} &=&\frac{g^{2}C_{F}}{4\pi ^{4}}\ C_{f}.
\end{eqnarray*}

The evaluation of \ various traces in the first two terms of (\ref{veff})
gives the following integral expression for this approximation of the CJT
effective potential:
\begin{eqnarray}
V[G_{g},G_{q}] &=&-i\int \frac{dp}{(2\pi )^{4}}\sum_{f}tr[\log [p_{\mu
}\gamma ^{\mu }\frac{1}{p_{\mu }\gamma ^{\mu }(1-\frac{C_{F}}{3}\frac{%
m_{g}^{2}}{p^{2}})+\frac{S_{f}}{p^{2}}}]-  \nonumber \\
&&p_{\mu }\gamma ^{\mu }\frac{1}{p_{\mu }\gamma ^{\mu }(1-\frac{C_{F}}{3}%
m_{g}^{2})+\frac{S_{f}}{p^{2}}}]+ \\
&&\frac{i}{2}\int \frac{dp}{(2\pi )^{4}}3N\ \left[ \log [\frac{p^{2}}{%
p^{2}+m_{g}^{2}}]-\frac{p^{2}}{p^{2}+m_{g}^{2}}\right],  \nonumber
\end{eqnarray}
in which the only trace remaining to be evaluated is the spinor one. This
relation directly shows that the fermion contribution is the sum of
identical functions of the quark parameters $S_{f}$.  Therefore, if the
expression has a minimum outside the value zero of these parameters, it
should be in a point in which all these \ constants take identical values.
This is the structure that had been predicted in Ref. \cite{colwitt} from
an analysis based on the large number of colours limit.  In that work the
central consideration was given to the plausibility that only the first loop fermion
correction should play a relevant role in the standard effective action of
the system. However, in the presence of ground state or vacuum
instabilities, the gap
equations (which determine the real spectrum and ground state of the system and the
possible field condensates) higher than one, and in particular the two
loop fermion contributions, should play an important role. This can be the case 
 because they are closely
linked with the Bethe--Salpeter bound state equations. Therefore, diagrams
contributing to the $V^{(2)}$ term of the effective action, 
such as the ones illustrated in 
 Fig. 2, can be expected to be relevant to the problem under
consideration. 

\begin{figure}
\begin{center}
\epsfig{figure=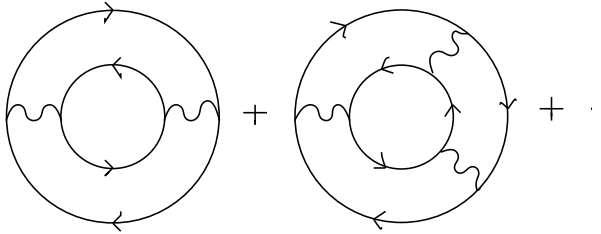,width=8cm}
\end{center}
\caption{Types of two fermions loop graphs contributing to $V^{(2)}$ 
which could  be relevant for the determination of the pattern 
of spontaneous flavor symmetry breaking in massless 
QCD}
\label{Fig.1}
\end{figure}
In another way, it is natural that this can be the situation in massless QCD, 
because the attractive binding 
forces are strong (unlike the case of the BCS 
superconductivity, where the attractive phonon
force works against the coulomb repulsion between electrons). In
QCD, as the quark-antiquark pairs have attractive and strong colour forces,
it would be expected that their condensates be spontaneously
generated.   But, as can be observed from the illustrated diagrams,
the participation of different quark loops can introduce contributions that
do not reduce to the sum of identical functions of each one of the parameters 
$S_{f}$. 
This fact seems to open the possibility for a dynamical generation
of condensates not having equivalent values for all the quarks. As within the
modified expansion dicussed here, the equality of all the condensates
(within the considered simple approximation) is equivalent to the equality
of the quark masses, the suggested possibilities for the spontaneous
generation of different quark condensates, would allow also the
spontaneous generation of `Democratic' quark mass matrices, as proposed in \cite
{fritzsch}. If these remarks are appropriate, the question about
the possible magnitudes of the condensates remains open as well.  These
issues should be considered in future extensions of this work.

\bigskip
\bigskip \noindent {\bf ACKNOWLEDGEMENTS }

I would like to express my gratitude for their helpful remarks and
conversations to F. Quevedo, M. Ruiz-Altaba, T.T. Wu, A. Schwartz, J.
Ellis, M. L\"uscher,\ J. A. Rubio, M. Mangano, P. Hern\'andez, S.
Pe\~naranda, G. Veneziano, K. Tamvakis, G. Altarelli, R.K. Ellis and B. Webber
during the visit in the Theory Division of CERN
where this work was done. I \ also strongly acknowledge the support of the
Associateship Scheme of the AS ICTP\ and the Theory Division of CERN, for
funding the travel from Havana \ and \ the visit to CERN, respectively, as
well as for the kind hospitality of the two related centres. The discussions
about the idea of the work with my colleague A. Gonzalez are also very much
acknowledged. Finally, I am indebted to the help in the proof-reading of the 
manuscript of Susy Vascotto and the warm hospitality of the TH Division 
Secretariat.

\end{document}